# Title: VIPER: Vectorial Interferometric Polarimeter for Electric-field Reconstruction

**Authors:** Tiancheng Huo[1,2], Li Qi[1], Jason J. Chen[1,2], Yusi Miao[1,2], Zhongping Chen[1,2]*

**Affiliations:**

[1]Beckman Laser Institute and Medical Clinic, Irvine, CA 92612, USA.

[2]Department of Biomedical Engineering, University of California, Irvine, Irvine, CA 92612, USA.

*Correspondence to: z2chen@uci.edu

**Abstract:** Comprehending the unique characteristics of structured ultrafast optical pulses is essential to the physics, applications and instrumentations of ultrafast lasers. However, full-vectrorial characterizations of the pulses remain challenging because of their multiple degrees of freedom. In this work, we implement and demonstrate the vectorial interferometric polarimeter for electric-field reconstruction (VIPER) which, for the first time, is able to precisely determine the specific three-dimensional E-field at every spatial-temporal coordinate throughout the entire detection region. This metrology provides a powerful tool for advanced ultrafast optics and paves the way for design optimization of ultrafast optics by providing the full information of the pulses, facilitating its applications in optical physics research, pump-probe spectroscopy and laser-based manufacturing.

**Main Text:** An optical pulse is usually a complicated classical electromagnetic field that can be described by its electric $\mathbf{E}(x,y,z,t)$ part as a three-dimensional time-varying vector field. The multidimensional, and vectorial reality of the optical pulse suggests that it can be structured in the spatial (*1*), or spectral-temporal (*2*) domain, as well as in its polarization states (*3*) individually or, in most cases, cooperatively. Kindled by the revolutionary development of pulse-shaping techniques, the capability to tailor optical pulses have brought immense impact to the spectroscopy (*4*), microscopy (*5*), lightwave communications (*6*), ultrafast optical imaging (*7*), optical trapping and tweezing (*8*), particle acceleration (*9*), material processing (*10*), and quantum sciences (*11*, *12*). Comprehensive and multidimensional characterization of these complicated electromagnetic radiations, therefore, plays an ever-increasing role in not only the accurate generation, meticulous modulation, and adaptive optimization of the structured light, but also the basic investigation of light-matter interactions.

Established spectral analysis methods (*13–19*), can resolve the temporal structures and the spectral phases, but the spatial and polarization profiles cannot be fully acquired. Although Mitchell et al. has partially characterized the polarization properties of wideband shaped beams (*20*), the spectral phases and temporal information are not maintained. While attosecond metrology can reconstruct the vectorial optical field for the ultrashort pulses, the spatial structures are unresolvable (*21*, *22*). To comprehensively understand the fundamental vectorial nature of complex optical pulses, we implement and demonstrate a vectorial interferometric polarimeter for electric-field reconstruction, termed VIPER to precisely determine the specific three-dimensional E-field vector at every spatial-temporal point throughout the entire pulse.

VIPER simultaneously records the polarization sensitive, horizontal (H channel) and vertical (V channel) auto-correlation functions of the self-referenced E-field in ultra-high spatial and temporal resolution (**Fig. 1**A). The three components of the field, especially the z component which has not previously been reported in literature, will be calculated by using the geometric Fresnel equations in spectral domain.

The spectral amplitude and phase for the H and V channels, $\widetilde{E}_H^S(x, y, \omega)$ and $\widetilde{E}_V^S(x, y, \omega)$, respectively, can be calculated after applying the standard techniques for spatially resolved Fourier transform spectrometer. As shown in **Fig. 1**B, the beam splitting interface, plane $\Pi$, divides the incident light into the transmitted part, $\widetilde{E}_p^S \hat{\mathbf{e}}_\mathbf{p}$, and reflected part, $\widetilde{E}_s^S \hat{\mathbf{e}}_\mathbf{s}$ where $\hat{\mathbf{e}}_\mathbf{p}$ and $\hat{\mathbf{e}}_\mathbf{s}$ are base vectors determined by the normal of $\Pi$, $\hat{\mathbf{e}}_\mathbf{n}$, and the unit wave vector, $\hat{\mathbf{e}}_\mathbf{k}$, which is the normal of wavefront $\varphi$, superscript 'S' stands for 'sample', such that:

$$\begin{cases} \hat{\mathbf{e}}_\mathbf{s} = \dfrac{(\hat{\mathbf{e}}_\mathbf{n} \times \hat{\mathbf{e}}_\mathbf{k})}{|\hat{\mathbf{e}}_\mathbf{n} \times \hat{\mathbf{e}}_\mathbf{k}|} \\ \hat{\mathbf{e}}_\mathbf{p} = \dfrac{(\hat{\mathbf{e}}_\mathbf{s} \times \hat{\mathbf{e}}_\mathbf{k})}{|\hat{\mathbf{e}}_\mathbf{s} \times \hat{\mathbf{e}}_\mathbf{k}|} \\ \hat{\mathbf{e}}_\mathbf{k} = \dfrac{\nabla \varphi}{|\nabla \varphi|} \end{cases}$$

Therefore, $\widetilde{E}_p^S$ and $\widetilde{E}_s^S$ can be described as $\widetilde{E}_H^S/(\hat{\mathbf{e}}_\mathbf{p} \cdot \hat{\mathbf{e}}_\mathbf{x})$ and $\widetilde{E}_V^S/(\hat{\mathbf{e}}_\mathbf{s} \cdot \hat{\mathbf{e}}_\mathbf{y})$, respectively. Then, the x, y, and z components of the incident light can be written as:

$$\begin{cases} \widetilde{E}_x^S = \widetilde{E}_p^S \hat{\mathbf{e}}_\mathbf{p} \cdot \hat{\mathbf{e}}_\mathbf{x} + \widetilde{E}_s^S \hat{\mathbf{e}}_\mathbf{s} \cdot \hat{\mathbf{e}}_\mathbf{x} \\ \widetilde{E}_y^S = \widetilde{E}_p^S \hat{\mathbf{e}}_\mathbf{p} \cdot \hat{\mathbf{e}}_\mathbf{y} + \widetilde{E}_s^S \hat{\mathbf{e}}_\mathbf{s} \cdot \hat{\mathbf{e}}_\mathbf{y} \\ \widetilde{E}_z^S = \widetilde{E}_p^S \hat{\mathbf{e}}_\mathbf{p} \cdot \hat{\mathbf{e}}_\mathbf{z} + \widetilde{E}_s^S \hat{\mathbf{e}}_\mathbf{s} \cdot \hat{\mathbf{e}}_\mathbf{z} \end{cases}$$

The spatial-temporal E-field, $\mathbf{E}^S(x, y, t)$, can be calculated by the inverse Fourier transformation of $\widetilde{\mathbf{E}}^S(x, y, \omega) = [\widetilde{E}_x^S, \widetilde{E}_y^S, \widetilde{E}_z^S]^\mathbf{T}$, where '**T**' denotes transpose operator.

**Fig. 2** presents the VIPER based full vectorial characterizations of the optical pulse with south pole state of the HOP sphere with a topological charge $l = +1$ produced by an orbital angular momentum-carrying femtosecond laser developed in-house. The doughnut-shaped amplitude structures observed in **Fig. 2**A(i-iv) are well matched with the conventional characterizations of this sample beam. However, the z component (**Fig. 2**A(iv)), mainly generated by the small angle ($\eta_0 = 1.68°$) between the sample and reference beam, also demonstrates the similar structure but with $1\% \pm 0.2\%$ of the corresponding maximum of total amplitude. The background-cleaned spatial-spectral phases (**Fig. 2**B(i-iii)) show a delicate spiral structure that corresponds to the theoretical helicity of +1 at 1026.6 nm. **Fig. 2**C(i-iii) are the well-defined polarization ellipses corresponding to the pairs of $(\widetilde{E}_x, \widetilde{E}_y)$, $(\widetilde{E}_x, \widetilde{E}_z)$ and $(\widetilde{E}_z, \widetilde{E}_y)$ calculated from **Fig. 2**A and **Fig. 2**B, respectively. **Fig. 2**C(i) globally shows the right-handed elliptical polarization well-matched with the designed polarization structure of the input beam. Furthermore, **Fig. 2**C(ii-iii) demonstrate nearly perfect linear polarization because of the tiny value of the z component. The corresponding gray background pattern are the first component of Stokes parameters. **Fig. 2**D(i) (Video S1) and **Fig. 2**D(ii-iv) are the spatial-temporal amplitudes of the total and individual electric field components, respectively. The shown isosurfaces (cyan and green) shown are set at 0.3 and 0.5 of the total maximum amplitude, respectively. The slice shows the energy density distribution at 482.9 fs (plane $\Sigma$). Both the doughnut-shaped features in the spatial domain, and temporal structure with ~522.8 fs full width at half maximum (FWHM) time duration are revealed in **Fig.2D**. **Fig. 2**E(i) (Video S2) provides the 3D visualization of global vectorial structure. **Fig. 2**E(ii-iii) (Video S3-S4) the corresponding 5.1X and 8.5X zoomed of **Fig. 2**E(i) viewed at points $\alpha$ and $\beta$ indicated in **Fig. 2**E(i) with the angle of view $(0, \pi)$ and $(3\pi/2, \pi/2)$, respectively. The color and scale of the arrows denote the magnitude, and the direction corresponds to the direction of the E-field. Presenting the evolutions of the E-field versus time at the given spatial points, **Fig. 2**E(ii-iii) demonstrate the temporal modelated polarization states. **Fig. 2**E(iv) (Video S5) is the electric (red arrows) and magnetic (green arrows) fields projected on $\mathbf{\Sigma}$. The magnetic field is determined by using the geometric relationship between **E**, **B**, and $\hat{\mathbf{e}}_\mathbf{k}$. It is clearly shown through the directions of the electric or magnetic field that the phase varies across $\mathbf{\Sigma}$ produced by $\eta_0$ and the intrinsic phase of the incident pulse. The results presented in **Fig. 2** validate the capability of VIPER for obtaining comprehensive information of complex optical pulses by resolving the 3D electric and magnetic field vectors throughout the entire detection regine.

**Fig. 3** indicates another important practical circumstance that VIPER can multidimensionally probe the scattered optical pulses generated by inserting an optical cleaning tissue before the PBS with the original left-handed circular polarization and Hermite–Gaussian (0,0) mode carrying incident beam. **Fig. 3**A(i) is the polarization ellipses of the pair $(\tilde{E}_x, \tilde{E}_y)$, which shows the modulated left-handed circular polarization at 1031.9 nm. Similarly with **Fig. 2**C(ii-iii), **Fig. 3**A(ii-iii) demonstrate nearly perfect linear polarization because of the tiny value of the z component. the spatial-temporal amplitude of the total and individual E-field components and the energy density distribution at 196.3 fs (plane $\mathbf{\Omega}$) demonstrate the spatial domain speckle patterns. Modulated by the optical cleaning tissue, the amplitude characterization helps signify the complexity of the spatial-temporal configurations that reveals the scattering and absorbing properties of the medium. **Fig. 3**C(i) (Video S7) depicts the 3D vectorial structure of the entire pulse. **Fig. 3**C(ii-iii) (Video S8-S9), the 5.1X and 8.5X zoomed of **Fig. 3**C(i) observed at points $\alpha$ and $\beta$, as shown in **Fig. 3**C(i) with the angle of view $(0, \pi)$ and $(3\pi/2, \pi/2)$, respectively. In **Fig. 3**C(iv) (Video S10), the electric (red arrows) and magnetic (green arrows) fields projected on $\mathbf{\Omega}$ demonstrate the polarizing sensitive modulation, by the scattering medium, on the optical properties of the incident pulse. By comprehensively and vectorially characterizing the ultrafast scattering fields, VIPER provides a promising foundation for investigating the optical properties of the matters as well as the pump-probe spectroscopy.

We have introduced the initial concept and the first implementation of VIPER for arbitrary input ultrafast optical patterns. The proposed methodology can completely manifest the full characterization of the spatial-temporal amplitudes of the three components of an ultrafast optical radiation carrying a complex polarization state through the whole detection region. It is proved to be a substantial tool for advanced ultrafast optics and paves the way for designing optimization of ultrafast optics by providing the full information of the pulses, facilitating its applications in optical physics research, pump-probe spectroscopy and laser-based manufacturing.

**Acknowledgments: Funding:** National Institutes of Health (R01HL-125084, R01HL-127271, R01EY-026091, R01EY-028662, and P41EB-01890); Air Force Office of Scientific Research (FA9550-17-1-0193); C.J. was supported by the National Science Foundation (DGE-1839285); In addition, the authors would like to thank Jing-Gao Zheng, PhD and Buyun Zhang, PhD for their meaningful discussion. **Author contributions:** T.H., L.Q. and Z.C. conceived and designed the research. Z.C. supervised the project. T.H. performed the experiments, data processing, numerical models, and numerical analysis. J.C. and Y.M. performed the simulation. L.Q., J.C., Y.M., and Z.Z. assisted in experiments. All the authors discussed the results and prepared the manuscript; **Competing interests:** Authors declare no competing interests; and **Data and materials availability:** All data is available in the main text or the supplementary materials.

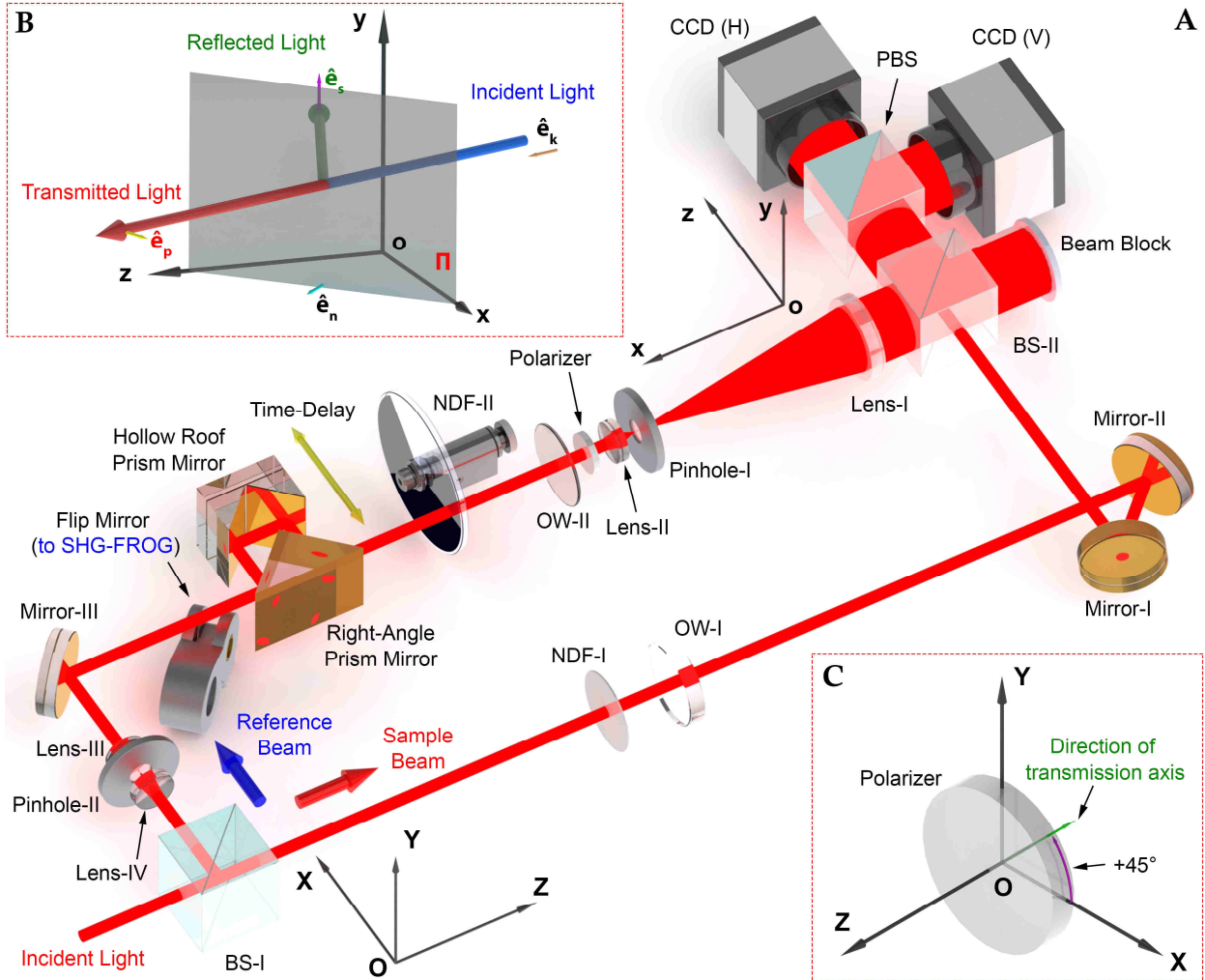

**Fig. 1. Vectorial interferometric polarimeter for electric-field reconstruction (VIPER).** (**A**) Schematic diagram of the VIPER. BS = beamsplitter cube; PBS = polarizing beamsplitter cube; OW = optical window; NDF = neutral density filter, SHG-FROG = the second-harmonic generation frequency-resolved optical gating. The red and blue arrows respectively denote the direction of the sample and reference beam. The global (associated with the laboratory) and local (with the PBS) reference systems are denoted by OXYZ and oxyz, respectively. (**B**) Geometry of the beam splitting interface (plane $\Pi$, gray surface) of the PBS. The yellow, pink, cyan and chocolate arrows denote the bases for p, s polarizations, normal of $\Pi$ and unit wavevector for the incident light, respectively. (**C**) Geometry of the polarizer. The directed angle between the transmission axis of the polarizer and the X-axis is usually set to $+45°$.

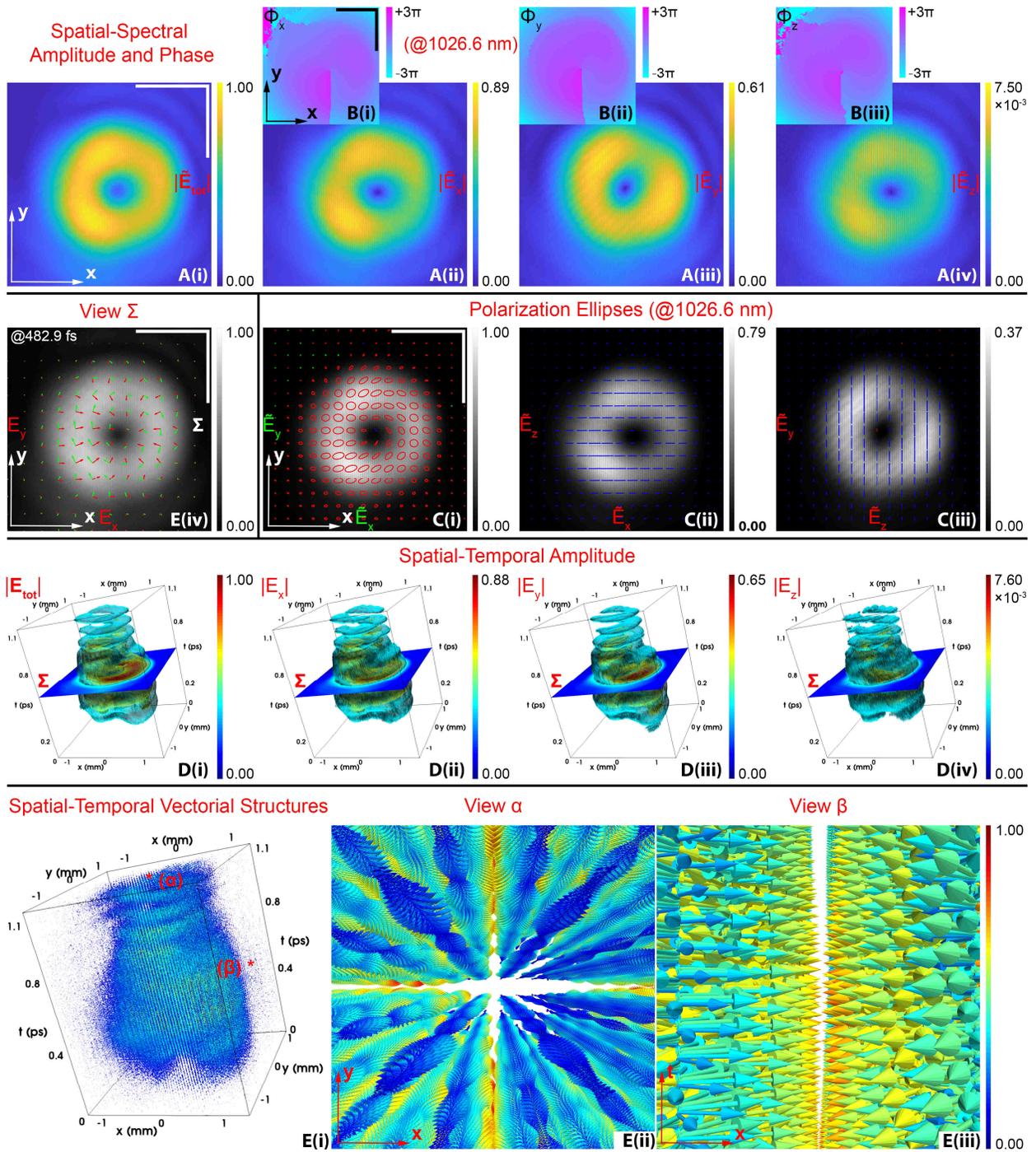

**Fig. 2. Full vectorial characterizations of the optical pulse with south pole state of the higher-order Poincaré (HOP) sphere (topological charge $l = +1$).** (**A**) and (**B**) the spatial-spectral amplitude of the total and individual E-field components, as well as the corresponding phase at 1026.6 nm. (**C**) the polarization ellipses and the first component of Stokes parameters for the pairs $(\tilde{E}_x, \tilde{E}_y)$, $(\tilde{E}_x, \tilde{E}_z)$ and $(\tilde{E}_z, \tilde{E}_y)$ at 1026.6 nm, respectively. The green line denotes the left-handed polarization, as the red to right-handed polarization, and blue to linear polarization. (**D**) the spatial-temporal amplitude of the total and individual E-field components. The isosurfaces (cyan and green) shown are set at 0.3 and 0.5 of the total maximum amplitude,

respectively. The slice shows the energy density distribution at 482.9 fs (plane **Σ**). (**E**) (**i**) the 3D global vectorial structure. (**ii-iii**) the 5.1X and 8.5X zoomed figures of (**i**) observed at point **α** and **β** (shown in (**i**)) with the angle of view $(0, \pi)$ and $(3\pi/2, \pi/2)$, respectively. The color and scale of the arrows denote the magnitude, and the direction corresponds to the direction of the E-field. (**iv**) the electric (red arrows) and magnetic (green arrows) fields projected on **Σ** with background of the corresponding energy density distribution. Scale bars represent 1 mm.

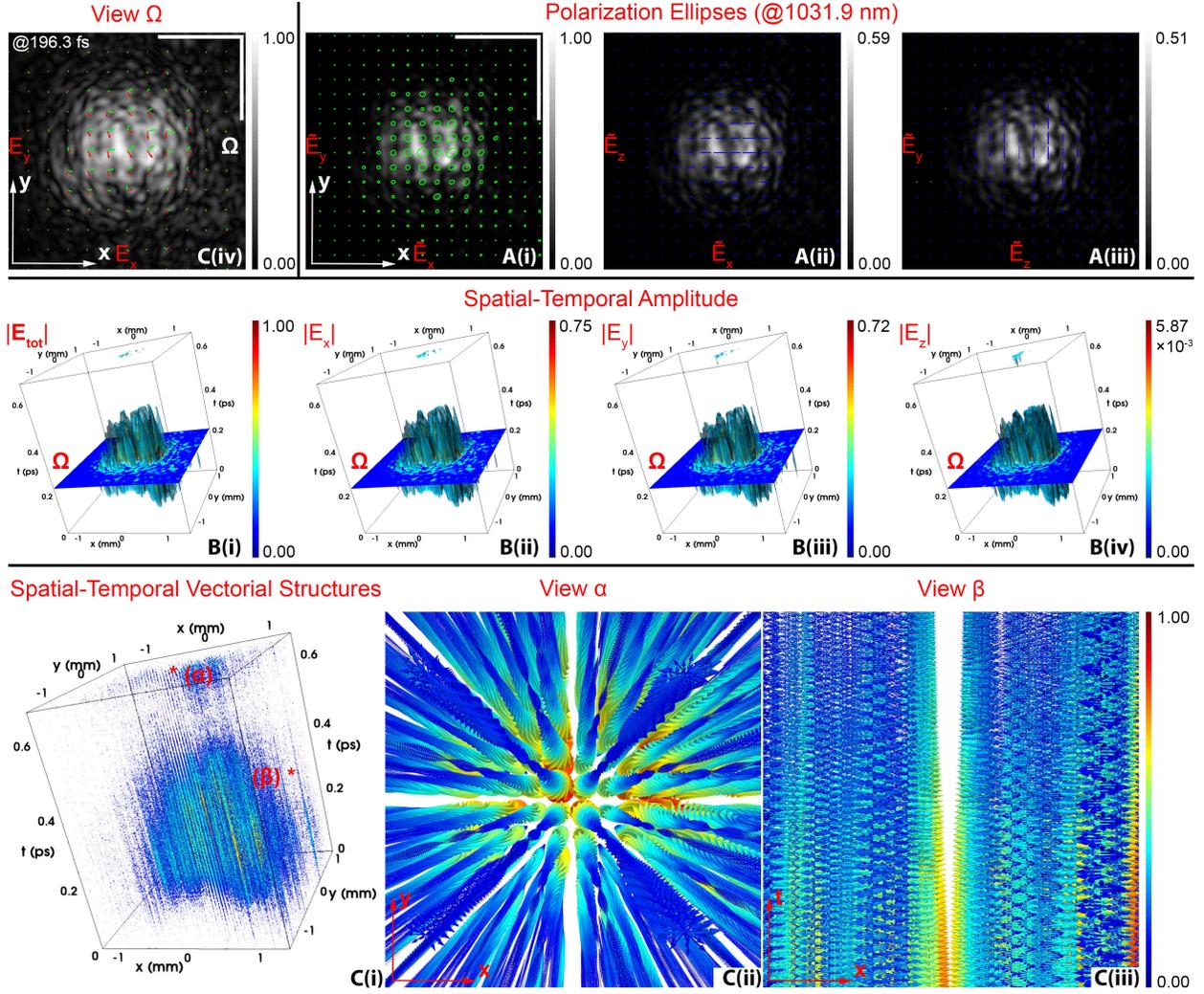

**Fig. 3. Full vectorial characterizations of the scattered optical pulse.** (**A**) the polarization ellipses and the first component of Stokes parameters for the pairs $(\tilde{E}_x, \tilde{E}_y)$, $(\tilde{E}_x, \tilde{E}_z)$ and $(\tilde{E}_z, \tilde{E}_y)$ at 1031.9 nm, respectively. (**B**) the spatial-temporal amplitude of the total and individual E-field components. The isosurfaces (cyan and green) shown are set at 0.3 and 0.5 of the total maximum amplitude, respectively. The slice shows the energy density distribution at 196.3 fs (plane **Ω**). (**C**) (**i**) the 3D global vectorial structure. (**ii-iii**) the 5.1X and 8.5X zoomed figures of (**i**) observed at point **α** and **β** (shown in (**i**)) with the angle of view $(0, \pi)$ and $(3\pi/2, \pi/2)$, respectively. The color and scale of the arrows denote the magnitude, and the direction corresponds to the direction of the E-field. (**iv**) the electric (red arrows) and magnetic (green arrows) fields projected on **Ω** with background of the corresponding energy density distribution. Scale bars represent 1 mm.